\documentclass{aa}
\usepackage{bm}
\usepackage{graphicx}
\usepackage{hyperref}
\usepackage{multirow}

\usepackage[varg]{txfonts}
\usepackage{natbib}
\usepackage{gensymb}

\newcommand{\be}{\begin{equation}}
\newcommand{\ee}{\end{equation}}
\newcommand{\rmd}{{\rm d}}

\newcommand{\msun}{{M}_{\sun}}

\bibpunct{(}{)}{;}{a}{}{,} 

\DeclareUnicodeCharacter{00A0}{ }

\begin{document}

\title{Polarized radiation from an accretion shock in accreting millisecond pulsars using exact Compton scattering formalism}

\titlerunning{Polarized radiation from accreting millisecond pulsars}

\authorrunning{A. Bobrikova et al.}

\author{Anna~Bobrikova\inst{1}
\and Vladislav Loktev\inst{1}
\and Tuomo Salmi \inst{2,1}
\and Juri~Poutanen\inst{1}}

\institute{Tuorla Observatory, Department of Physics and Astronomy, 20014 University of Turku, Finland \\ \email{anna.a.bobrikova@utu.fi} 
\and Anton Pannekoek Institute for Astronomy, University of Amsterdam, Science Park 904, 1098XH Amsterdam, the Netherlands
}

\date{Received 8 May 2023 / Accepted 11 August 2023}

\abstract{
Pulse profiles of accreting millisecond pulsars can be used to determine neutron star (NS) parameters, such as their masses and radii, and therefore provide constraints on the equation of state of cold dense matter. 
Information obtained by the \textit{Imaging X-ray Polarimetry Explorer} (\textit{IXPE}) can be used to decipher pulsar inclination and magnetic obliquity, providing ever tighter constraints on other parameters.  
In this paper, we develop a new emission model for accretion-powered millisecond pulsars based on thermal Comptonization in an accretion shock above the NS surface. 
The shock structure was approximated by an isothermal plane-parallel slab and the Stokes parameters of the emergent radiation were computed as a function of the zenith angle and energy for different values of the electron temperature, the Thomson optical depth of the slab, and the temperature of the seed blackbody photons. 
We show that our Compton scattering model leads to a significantly lower polarization degree of the emitted radiation compared to the previously used Thomson scattering model.  
We computed a large grid of shock models, which can be combined with pulse profile modeling techniques both with and without polarization included.  
In this work, we used the relativistic rotating vector model for the oblate NS in order to produce the observed Stokes parameters as a function of the pulsar phase. 
Furthermore, we simulated the data to be produced by \textit{IXPE} and obtained constraints on model parameters using nested sampling. 
The developed methods can also be used in the analysis of the data from future satellites, such as the \textit{enhanced X-ray Timing and Polarimetry} mission. 
}


\maketitle

\section{Introduction}\label{intro}

Rapidly rotating accreting neutron stars (NSs) can be used to study the properties of the extremely dense matter inside their cores. 
Some of these are accretion-powered millisecond pulsars (AMPs), which show X-ray pulsations at the stellar spin frequency. Their emission profiles can be modeled, accounting for light bending and other relativistic effects, and from these profiles, the information about the mass and radius of the NS can be extracted\citep[see e.g.,][]{PFC83,ML98,PG03,PB06,MLC07,lomiller13,2016RvMP...88b1001W,SNP18,BLM_nicer19}. 
The mass and radius constraints can be translated to the equation of state (EOS) constraints for matter inside the NS core \citep[e.g., see][]{lindblom1992,Lattimer12ARNPS,Hebeler2013,Baym2018}. 

This technique was recently applied to the case of rotation-powered millisecond pulsars using data from the Neutron Star Interior Composition Explorer \citep[NICER,][]{MLD_nicer19,RWB_nicer19,Miller2021,Riley2021,Salmi2022}. 
In these sources, the hot regions on the NS surface are heated by a magnetospheric return current and the emission escaping the NS surface can be computed using self-consistent atmosphere models \citep[see e.g.,][]{ZPS96,Ho01,HR06,HH09,Haakonsen2012,Salmi20}. 
For AMPs, these models are not adequate because the emitted radiation is Comptonized in an accretion ``shock'' formed above the hot region. 
In reality, a standard gasdynamical shock might not exist at all, but the kinetic energy of incoming particles is released in the surface layer through Coulomb collisions and plasma instabilities  \citep{Zeldovich1969,SPW18,Salmi20}.   
Comptonization of soft photons generated within the layer as well as by the underlying NS surface is then responsible for the power-law spectra extending up to 100 keV as observed from AMPs \citep{PG03,GP05,Falanga05a,Falanga05b,Falanga07,Falanga11,Falanga12}.  
Electron scattering also causes the radiation to be significantly polarized. 
Variations of polarization with pulsar phase can be used to determine geometrical parameters such as inclination angle and magnetic obliquity \citep{VP04}, which in their turn allow us to improve constraints on the NS mass and radius.  

Previously, polarized emission models for AMPs used the Thomson scattering approximation in an optically thin NS atmosphere \citep[see][]{ST85,VP04,SLK21}. 
However, the polarization degree due to Compton scattering on hot electrons  strongly depends on the electron temperature \citep{NP94a,Pou94ApJS}; for example, for 100~keV electrons, polarization is less than 50\% of that for Thomson scattering.
On the other hand, the models for nonpolarized radiation ---in the context of pulse profile modeling--- typically employ approximate formulas for the anisotropy of the radiation and empirical models for the Comptonized spectra \citep[see e.g.,][]{PG03,LMC08,Steiner2009,SNP18}.  
Self-consistent accretion-heated atmosphere models have also been developed \citep[see e.g.,][]{zampieri1995,deufel2001,SPW18}, although these are relatively computationally expensive and have a large number of free parameters. 

In this work, we applied the formalism for Compton scattering in a hot slab \citep{NP93,PS96} to compute the Stokes parameters of the emergent radiation as a function of energy and emission angle using only three model parameters: the electron temperature and optical depth of the hot slab on top of the NS surface, and the temperature of the seed blackbody photons coming from the NS. 
By employing pulse profile modeling for polarized radiation from rapidly rotating oblate NSs \citep{poutanen20,LSNP20}, we also simulated data for the \textit{Imaging X-ray Polarimeter Explorer} \citep[\textit{IXPE},][]{Weisskopf2022} and updated the NS geometry parameter constraints previously predicted in \citet{SLK21} using the Thomson scattering model. 
These methods can also be applied when analyzing the observations from \textit{IXPE}\footnote{\textit{IXPE} was launched in December 2021, but has not yet been able to observe an AMP in outburst.} or from future X-ray polarimetric missions such as the \textit{enhanced X-ray Timing and Polarimetry} mission \citep[\textit{eXTP};][]{Zhang19,dmatter_extp}. 
The atmosphere look-up tables produced here can also be combined with any AMP pulse profile modeling codes.\footnote{The tables are available at \url{https://github.com/AnnaBobrikova/ComptonSlabTables}}.

The remainder of this paper is structured as follows.
In Sect.~\ref{sec:local}, we present the theory that describes the formation of polarized radiation in a Comptonizing slab above the NS surface. 
We then apply this theory to obtain the intensity and polarization of the radiation escaping from the slab. 
We then describe the method to obtain pulse profiles using our new radiation model. 
We compare our resulting spectra and pulse profiles to those obtained using previous models. 
We also discuss pulse profiles obtained for different NS parameters.
In Sect.~\ref{sec:simul}, we generate synthetic data and apply our fitting routine to determine the NS parameters from the data. 
We discuss the applications of our model and predictions for the upcoming observations in Sect.~\ref{sec:discussion}. 
In Sect.~\ref{sec:summary}, we conclude and summarize our findings.

\section{Emission model}\label{sec:local}

\subsection{Radiative transfer equation}\label{sec:emission_model}

We consider a simple NS atmosphere model in which the atmosphere is a plane parallel slab consisting of electrons and lying above an optically thick source that radiates as a  blackbody.
The hot slab has the Thomson optical depth $\tau_{\mathrm{T}}=\sigma_{\mathrm{T}} n_{\mathrm{e}} H$, where $H$ is the vertical height, $\sigma_{\mathrm{T}}$ is the Thomson cross section, and $n_{\mathrm{e}}$ is the electron concentration.
The electron gas in the atmosphere is considered to be isotropic and isothermal with the electron temperature $T_{\mathrm{e}}$.
The momentum distribution of electrons is given by relativistic Maxwellian distribution characterized by the dimensionless temperature $\Theta_{\mathrm{e}}=kT_{\rm e}/m_{\rm e}c^2$:
\be 
\label{eq:Maxwell}
f_{\mathrm{M}}(\gamma)= \frac{e^{-\gamma /\Theta_{\mathrm{e}}}}{4\pi \Theta_{\mathrm{e}} \ K_2(1/\Theta_{\mathrm{e}})} ,
\ee
where $\gamma$ is the electron Lorentz factor and $K_2$ is the modified Bessel function of the second kind.  
This distribution is normalized to unity: 
\be 
\label{eq:Maxwell-norm}
\int_0^\infty f_{\mathrm{M}}(\gamma)  \rmd^3 p = 4\pi \int_0^\infty f_{\mathrm{M}}(\gamma)  p^2 \rmd p = 1 , 
\ee
where $p=\sqrt{\gamma^2-1}$ is the electron dimensionless momentum. 
The radiation field can be described by the Stokes vector $\bm{I}(\tau,x,\mu)$, where $\mu$ is the cosine of the zenith angle, an angle between the normal to the slab and the direction of photon propagation, and $x=E/m_{\rm e}c^2$ is the photon energy measured in the units of the electron rest mass.  

The Stokes vector usually contains four components $I$, $Q$, $U$, and $V$, but because of the azimuthal symmetry and the absence of sources of circular polarization, the latter two are equal to zero. 
We therefore use just two Stokes parameters $\bm{I}(\tau,x,\mu)= (I,Q)^{\rm T}$, where superscript T denotes the transposed vector. 

The photon distribution at the bottom of the slab is considered to be Planckian of the temperature $T_{\mathrm{bb}}$. 
The incident radiation (for zeroth scattering order, $n=0$) at the bottom of the slab is given by the Stokes vector  
\be
\vec{I}_{n=0} (\tau=0,x,\mu) = 
\frac{2m_{\mathrm{e}}^4c^6}{h^3}\frac{x^3}{e^{x/\Theta_{\mathrm{bb}}}-1} 
\left(
\begin{array}{c}  
1 \\ 0 
\end{array} 
\right) ,
\ee   
where $\Theta_{\mathrm{bb}}=kT_{\mathrm{bb}}/m_{\rm e}c^2$.  
In order to describe the propagation of polarized radiation through the hot electron slab, we solve the radiative transfer equation (RTE)  \citep{NP94a,PS96}: 
\begin{equation}
\mu \frac{\rmd \vec{I}(\tau,x,\mu)}{\rmd\tau} = 
- \sigma(x)  \vec{I}(\tau,x,\mu) + \vec{S}(\tau,x,\mu) ,
\end{equation}
where $\rmd\tau=\sigma_{\rm T}n_{\rm e} dz$ is the Thomson optical depth, $\vec{S}$ is the source function (also a Stokes vector), and  $\sigma(x)$ is the dimensionless Compton scattering cross-section (in units of the Thomson cross-section $\sigma_{\rm T}$).  
We solve the RTE using the iterative scattering method of \citet{PS96}, where the intensity is represented as a series expansion in scattering orders. 
The Stokes vector of unscattered radiation at all optical depths is 
\begin{equation}
\vec{I}_{n=0} (\tau,x,\mu) = \vec{I}_{n=0} (0,x,\mu) \ 
{\rm e}^{-\sigma(x) \tau/\mu}. 
\end{equation} 
From the known Stokes vector at the $n$-th scattering order, we find the source function for the following scattering order as 
\begin{equation} \label{eq:cs_source}
\vec{S}_{n+1}(\tau,x,\mu)  
= x^{2}\!\! \int_0^{\infty}\!\! \frac{\rmd x_1}{x_1^{2}} \int_{-1}^1\!\!\! \rmd \mu_1 \ 
\hat{\vec{R}} (x,\mu; x_1,\mu_1)  
\vec{I}_{n}(\tau,x_1,\mu_1) ,
\end{equation}
where $\hat{\vec{R}}$ is the $2\times2$ azimuth averaged redistribution matrix describing Compton scattering by isotropic  electrons \citep[see Appendix A1 in][]{PS96}. 
If the source function is known, the Stokes vector is found via the formal solution of the RTE: 
\begin{equation} \label{eq:formal_rte}
\vec{I}_{n+1}(\tau,x,\mu) \! =\!
\left\{ 
\begin{array}{ll}  
\displaystyle 
\!\!\int_{0}^{\tau} \!\! \frac{\rmd \tau'}{\mu} \ \vec{S}_{n+1}(\tau',x,\mu) \  {\rm e}^{-(\tau-\tau')\sigma(x)/\mu}, & \mu>0, \\ 
& \\
\displaystyle \!\!\int_{\tau}^{\tau_{\rm T}}  \!\!\frac{\rmd \tau'}{(-\mu)} \ \vec{S}_{n+1}(\tau',x,\mu) \ {\rm e}^{-(\tau'-\tau)\sigma(x)/(-\mu)}, & \mu<0 .
\end{array}
\right.
\end{equation}   
Iterations proceed until the required accuracy of the total Stokes vector  $\vec{I}=\sum_{n}\vec{I}_n$ is achieved so that the maximal contribution of the next scattering is less than $1\%$ of the total spectrum in all energies and angles. 

 \begin{figure*}
\centering
  \includegraphics[width=0.47\textwidth]{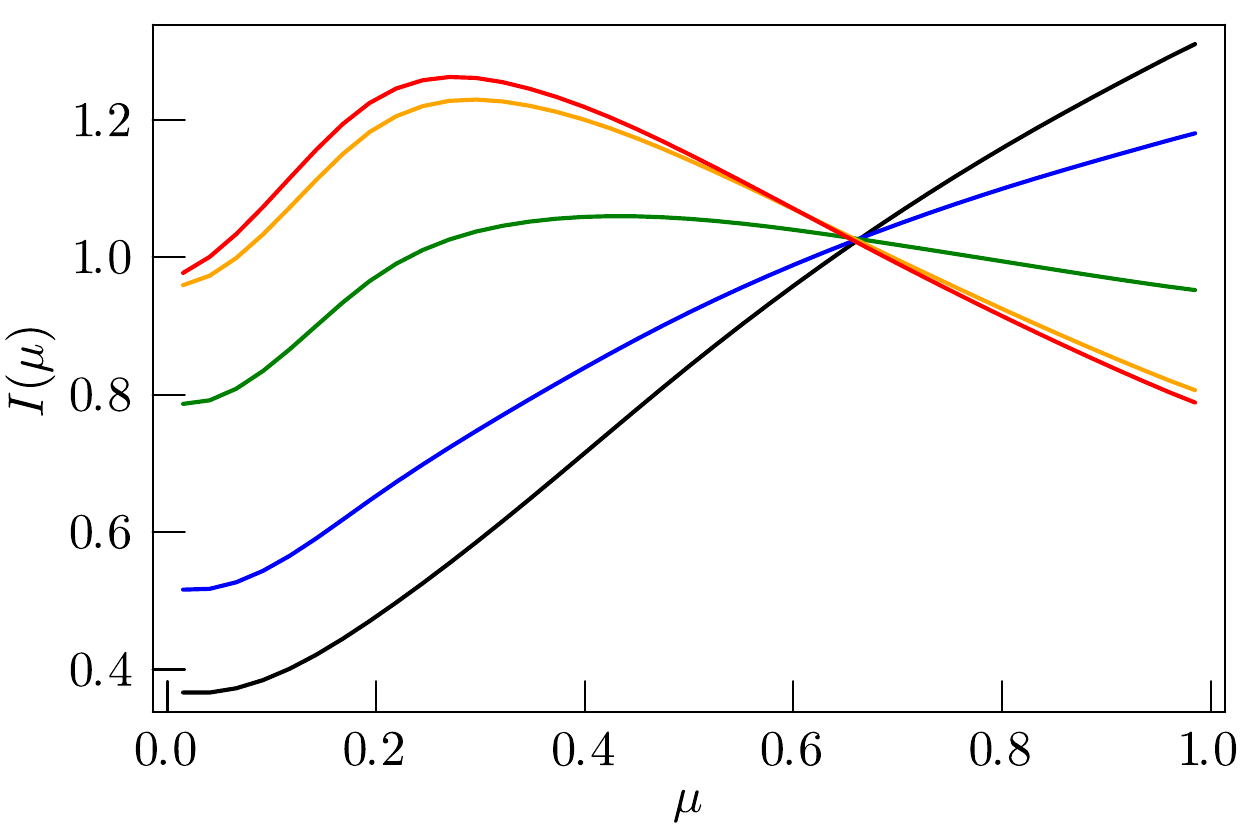}
  \includegraphics[width=0.47\textwidth]{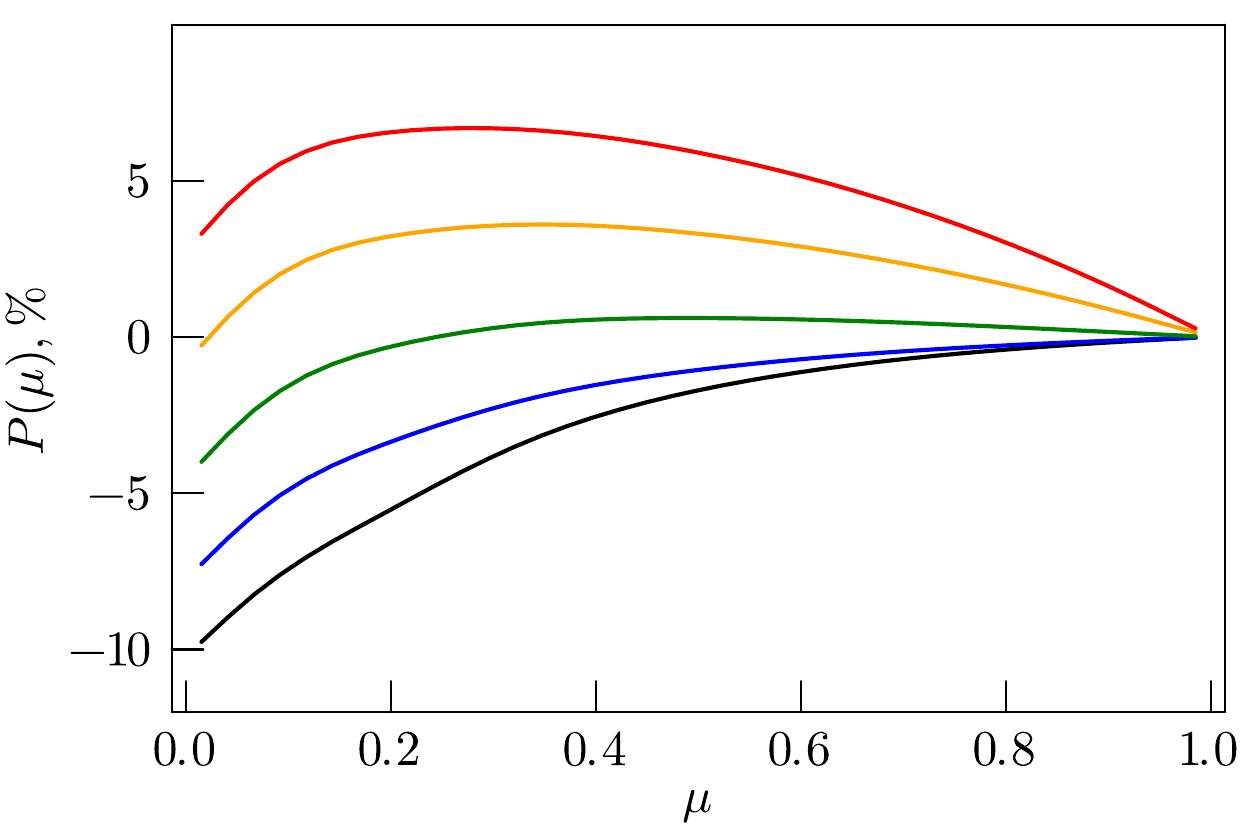}
\caption{Intensity of the emergent radiation (left) and the PD (right) as functions of the cosine of the zenith angle. 
The atmosphere parameters are  $\tau_{\mathrm{T}}$ = 1.0, electron temperature $T_{\rm e}$=50 keV, and seed photon temperature $T_{\rm bb}$=1 keV. 
Black, blue, green, orange, and red solid lines show the model for photon energies of 2, 5, 8, 12, and 18 keV, respectively. 
The angular dependence of the intensity is normalized so that $\int_{0}^{1}\mu I(\mu) \rmd \mu  = 1/2$.}
  \label{fig:Compton1}
\end{figure*}

 \begin{figure*}
  \centering
  \includegraphics[width=0.47\textwidth]{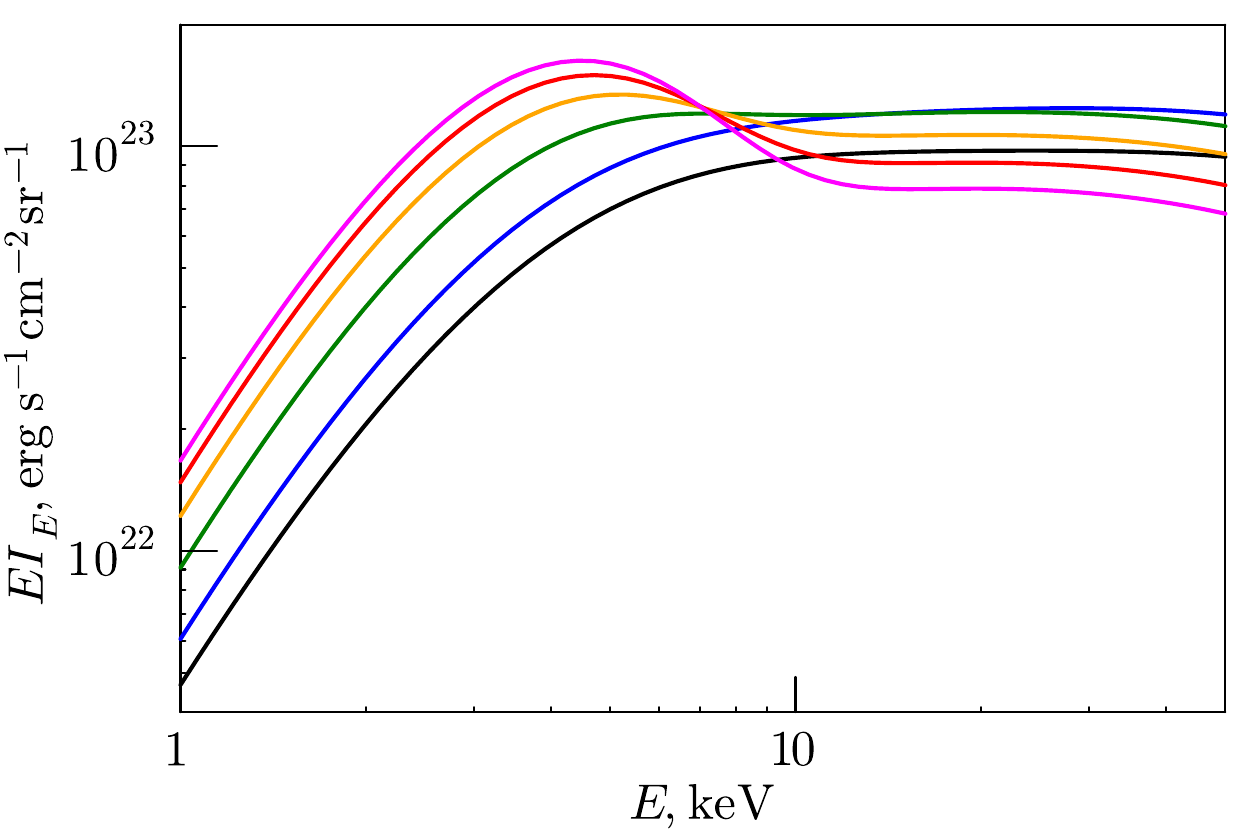}
  \includegraphics[width=0.47\textwidth]{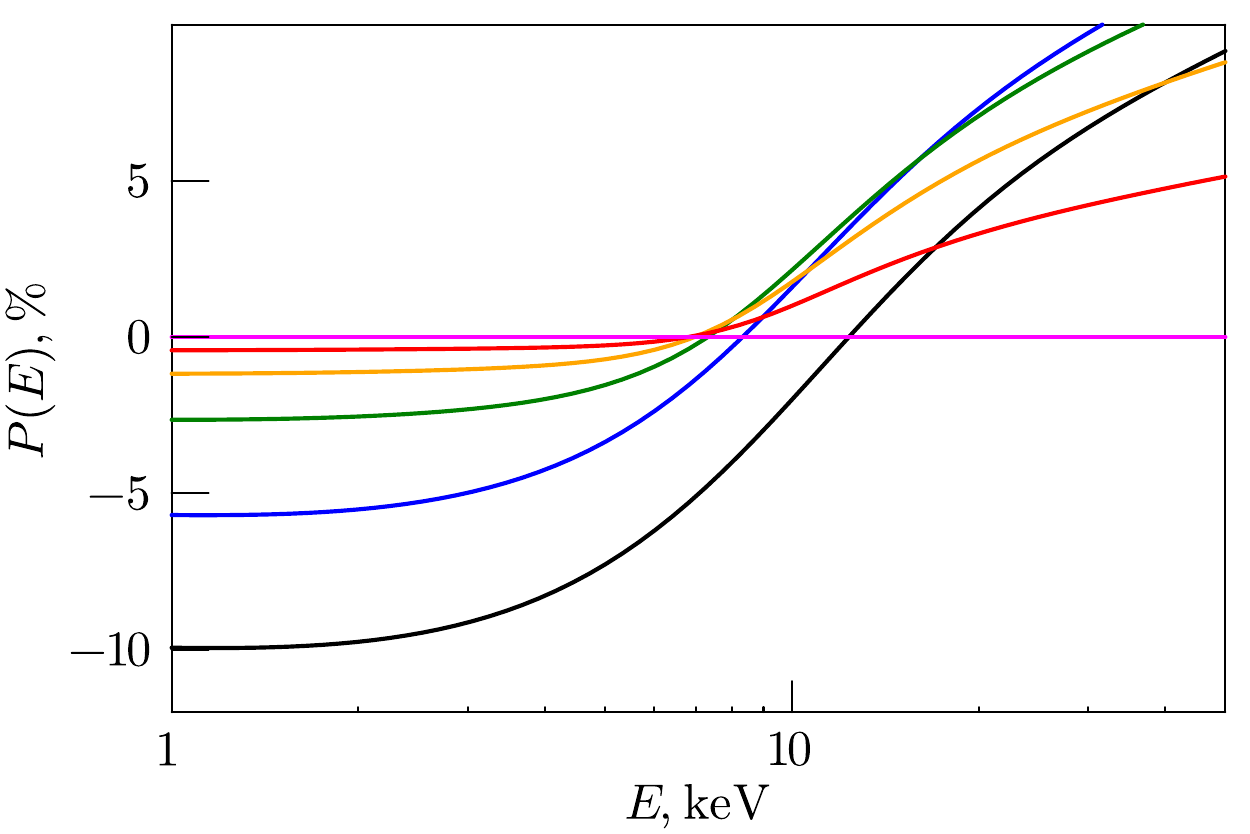}
\caption{Intensity of emergent radiation (left) and the PD (right) as functions of photon energy. 
The same atmosphere parameters are used as in Fig.~\ref{fig:Compton1}.
Black, blue, green, orange, red, and magenta solid lines show the model for $\mu$ = 0.0, 0.2, 0.4, 0.6, 0.8, and 1.0 respectively.  }
  \label{fig:Compton2}
\end{figure*}

Finally, we can compute the polarization degree (PD) of the emergent radiation as 
\be
\label{PD}
P(x,\mu)=100\%\frac{Q(\tau_{\mathrm{T}},x,\mu)}{I(\tau_{\mathrm{T}},x,\mu)}. 
\ee
The positive values of $P$ imply that the polarization vector lies in the meridional plane defined by the slab normal and the photon momentum, while for the negative values, the polarization vector is perpendicular to that plane, as in the case of optically thick electron scattering atmosphere \citep{Cha60,Sob63}.

\subsection{Spectro-polarimetric properties of accretion shocks} 
\label{sec:beaming}

We can now apply the derived formalism to predict the beaming patterns, energy spectra, and polarization of the accretion shock at the NS surface. 
We take as a fiducial set of  parameters: $\tau_{\mathrm{T}}= 1$, $T_{\mathrm{e}} = 50$~keV, $T_{\mathrm{bb}} = 1$~keV. 
Figures~\ref{fig:Compton1} and \ref{fig:Compton2} show the spectrum and the PD as functions of the cosine of the zenith angle and energy, respectively, for this set of parameters.
The parameters of the source are the same as those used in the Thomson model of \citet{SLK21} (see their Figs. 1 and 2), and so these figures can be used to illustrate the impact of using an exact Compton scattering description instead of the Thomson scattering model in the slab of hot electrons. 

When comparing our results with those of ~\citet{SLK21}, we note that the PD differs significantly while the spectrum remains mainly unchanged. 
This effect is most obviously seen in the angular dependence of the PD. 
The absolute values of the PD are reduced by a factor of two for the photon energies of 2, 5, and 8 keV, and increased at higher energies. 
For Compton scattering, the PD changes the sign already at the photon energy of 8 keV and is positive at higher values, while for Thomson scattering the sign changes at photon energies above 18 keV.
Last but not least, the maximum in the absolute value of the PD for the photon energies higher than 8 keV has shifted from $\mu = 0$ to higher values of the cosine of the zenith angle. 
Keeping in mind the energy range of the \textit{IXPE} satellite (2--8 keV), we can say that compared to the previously developed model for AMPs that used Thomson scattering, the Compton scattering model predicts a lower PD. This result is in agreement with our expectations, as for hot electrons the relativistic aberration becomes important \citep{Pou94ApJS}. In the case of cold electrons, the PD of the Thomson singly scattered radiation is highly dependent on the scattering angle, reaching 100\% at  90\degr. For relativistic electrons (but still in the Thomson scattering regime), the same happens in the electron rest frame; however, because of the relativistic motion of the electron, the photon in the external frame is preferentially scattered in the direction of the electron motion, resulting in zero final PD independent of the scattering angle \citep{BCS70,NP93}. 
In our case of 50~keV electrons, the electrons are not yet relativistic and the maximum PD is reduced to 60\% from the Thomson case. 
We can predict here that with the higher electron temperature, the PD will decrease further. This will be examined in the following section.

\subsection{Grid of models} \label{sec:grid}

 \begin{figure*}
  \centering
  \includegraphics[width=0.98\textwidth]{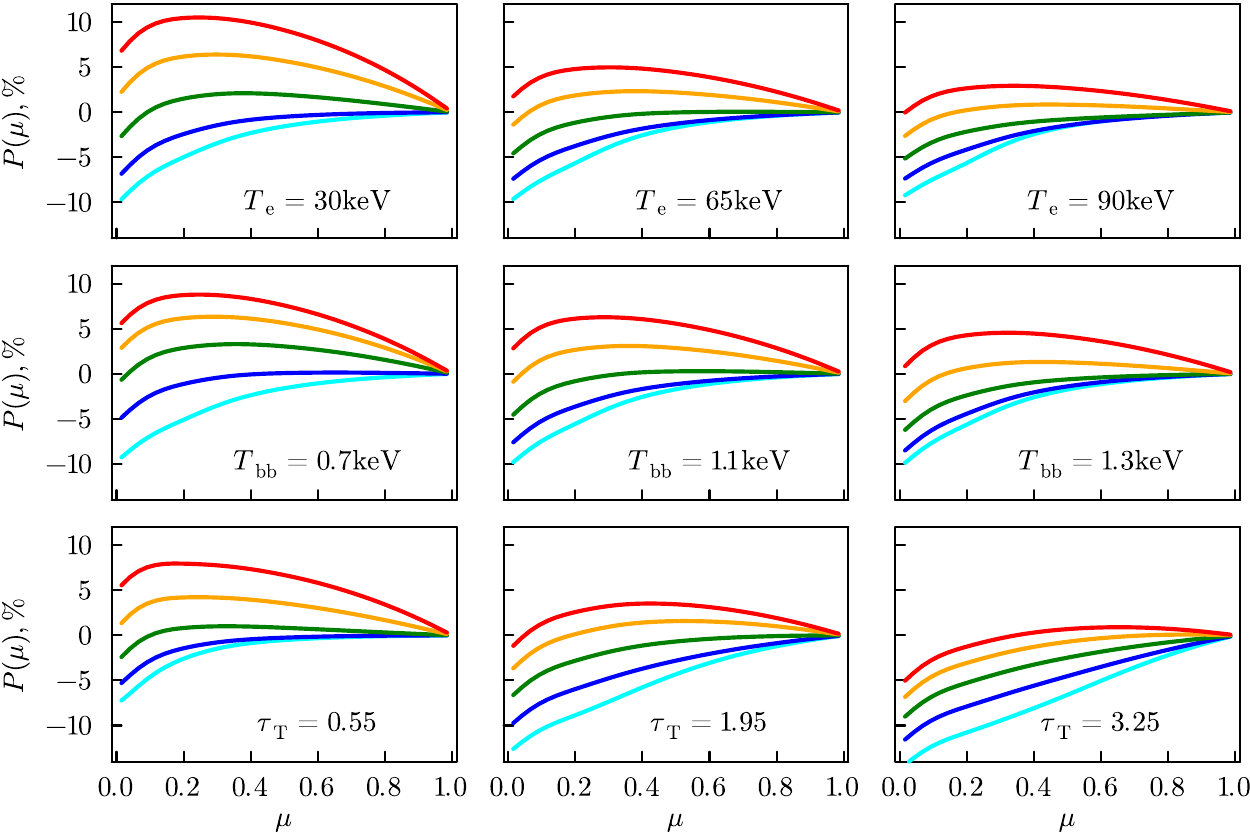}
\caption{Angular dependency of the PD for different atmosphere parameters. 
For each plot, we vary only one parameter (marked on the plot) while other parameters are from the fiducial set ($\tau_{\mathrm{T}}$ = 1, $T_{\mathrm{e}}$ = 50~keV, $T_{\mathrm{bb}}$ = 1~keV). 
  Cyan, blue, green, orange, and red solid lines show the PD on photon energies of 2, 5, 8, 12, and 18 keV, respectively. 
  }
  \label{fig:pmu}
\end{figure*}

The model we consider for the accretion shock has three parameters: 
the optical depth of the Comptonizing plasma $\tau_{\mathrm{T}}$, the dimensionless electron temperature $\Theta_{\rm e}$, and the dimensionless characteristic photon temperature $\Theta_{\rm bb}$ (associated with $T_{\rm e}$ and $T_{\rm bb}$, respectively). 
To determine these parameters from the upcoming \textit{IXPE} data, we need a dense grid of models. 
For the optical depth of the shock $\tau_{\mathrm{T}}$, we chose the lower and higher value of 0.5 and 3.5, respectively, and varied the parameter in steps of 0.1. 
For the electron temperature, we varied $\Theta_{\mathrm{e}}$ from 0.04 to 0.2 (i.e., $T_{\mathrm{e}}$ from $\sim$20 to 100 keV) with a step of 0.004 (i.e., $\sim$2 keV). 
For the seed photon temperature, the dimensionless $\Theta_{\mathrm{bb}}$ was varied in the interval (1--3)$\times10^{-3}$ (i.e., $T_{\mathrm{bb}}=$0.5--1.5~keV) in steps of $2\times10^{-4}$ (i.e., $\sim$0.1 keV). 
Altogether, we computed 13981 models. 

The impact of the NS atmosphere parameters on the PD can be seen in Fig.~\ref{fig:pmu}. For the electron temperature, we see the expected decrease in the PD with $T_{\rm e}$ as electrons become more relativistic. For the two other parameters, the main feature is that with increasing the parameter value, the change of sign of the PD happens at higher energies. Within the mentioned intervals of parameter values, the impact is most significant with increasing the optical depth of the atmosphere and least significant with changes to the seed photon energy. For the \textit{IXPE} observational range of 2--8 keV, this means that,  in general, PD increases with the increase in these parameters. 

In order to use the calculated Stokes parameters more effectively, we wrote a 3D interpolation routine over this grid of parameters, so that for any arbitrary combination of these parameters within the intervals mentioned, we can obtain the intensity and PD of the emission as functions of energy and zenith angle. We tested that the results of this interpolation match the direct calculation results for any combination of the parameters within the grid 
to within 8\% (within 1\% in the \textit{IXPE} energy range). 

\subsection{Pulse profiles and polarization}\label{sec:pulses}

We calculated the pulse profiles from the known beaming pattern and PD exactly as in \citet{SLK21}. 
After obtaining the beaming pattern and PD of the emission on top of the NS atmosphere, we transport them from the hot emitting regions of the star (called hot spots) to the observer following the formalism described in \citet{Morsink_2007}, \citet{AlGendy_2014}, \citet{poutanen20}, and \citet{LSNP20}. 
We consider an oblate, rapidly rotating star in the Schwarzschild metric (hence, the effects of rotation on space-time are neglected). 
Gravitational redshift and the light-bending effect, as well as the effect of the fast rotation of the star, are accounted for in the polarization angle (PA) transfer. The only difference in approach is the reference radius used in time-delay calculation. 
While
\citet{LSNP20} use the exact radius of the star at the center of the spot, we follow the path described in \citet{PB06} and applied in \citet{SLK21} and use the equatorial radius as a reference radius. 

\begin{figure*}
\centering
\includegraphics[width=0.8\textwidth]{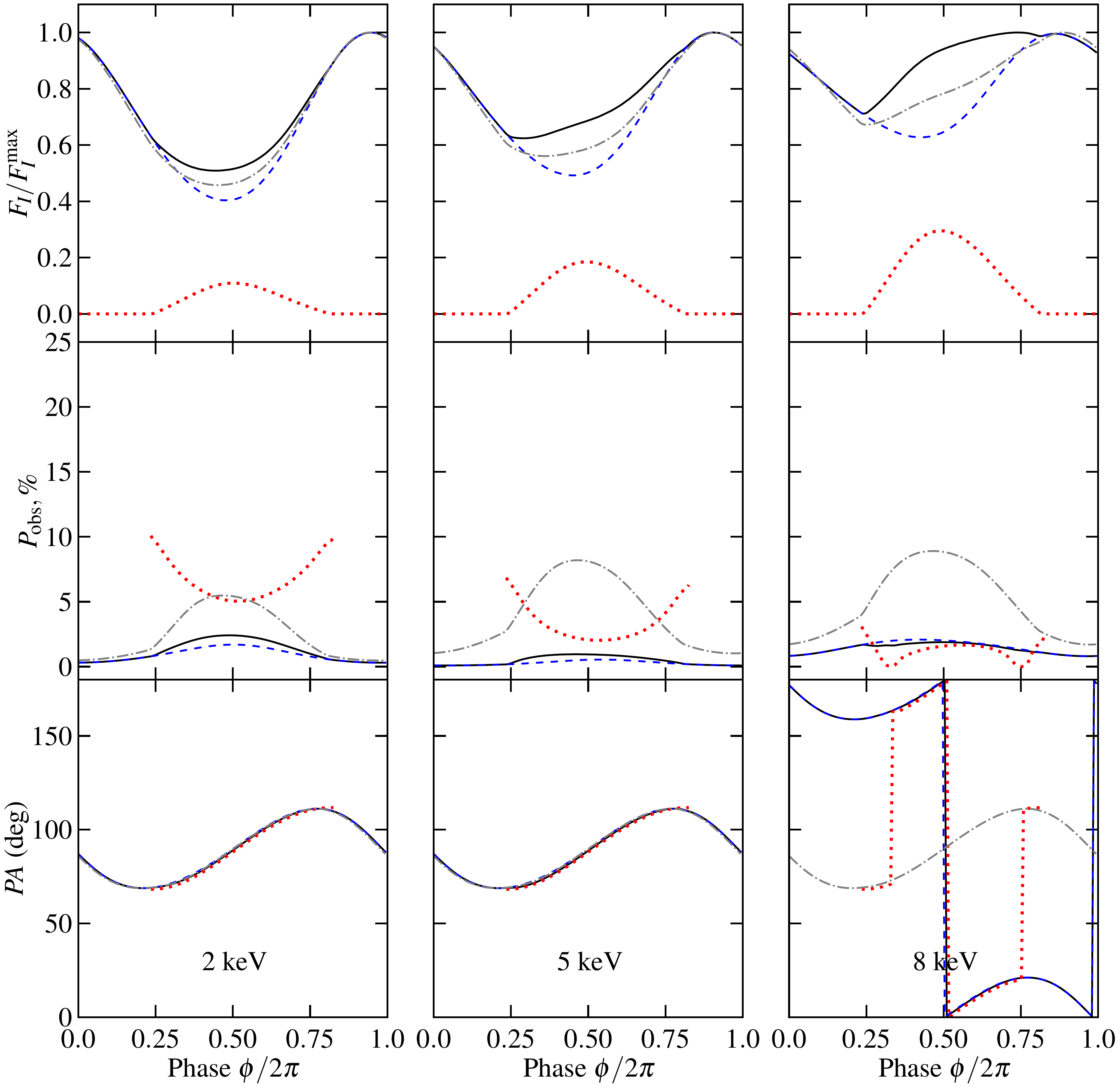}
\caption{Pulse profiles of the observed flux, PD, and PA for two antipodal spots shown for three different energies (2, 5, $8~\mathrm{keV}$). The solid black curves correspond to the total flux (PD, PA), the dashed blue lines correspond to the contribution from the primary spot, and the dotted red lines are for the secondary spot. The dash-dotted gray lines correspond to the Thomson model from \citet{SLK21} (combined from two spots). The NS parameters in both models are those shown in Table~1 of \citet{SLK21}.}
\label{fig:polpulse1}
\end{figure*}

We again return to the results from \citet{SLK21} and compare the pulse profiles coming from the model with Compton scattering in the atmosphere with previously obtained ones using Thomson scattering. Figure~\ref{fig:polpulse1} presents this comparison for the case of two antipodal spots, both combined and individually from each spot, for the NS with atmospheric parameters from the fiducial set  (inclination $i$ = 60\degr\ and magnetic co-latitude $\theta$ = 20\degr\ as in \citealt{SLK21}). 
We see that the flux does not change significantly, which is in agreement with the beaming patterns not changing either. The PD is now decreasing from 2 to 5~keV and then slowly increasing towards 8~keV. From inspection of the corresponding lines in Fig.~\ref{fig:Compton1}, and accounting for the gravitational redshift, it is clear that the PD is following the same pattern for the corresponding energies of the emitted photons. The `jump' by 90\degr\ in the PA for the 8~keV photons comes from the PD changing sign.

The main conclusion here is that we expect a significant drop in the observed PD compared to that predicted by \citet{SLK21}. The impact of this reduction in PD on the possibility of observing the AMPs is further discussed in Sects.~\ref{sec:simul} and ~\ref{sec:discussion}.

\section{Synthetic data and their analysis}\label{sec:simul}

\subsection{Generating the synthetic data}

The developed model can be further used to analyze the data. Polarimetric measurements of the AMPs are yet to be carried out. In the absence of observations, we used the \textsc{ixpeobssim} framework \citep{BALDINI2022101194} (version 16) with the \textsc{pcube} algorithm to generate several samples of synthetic data imitating the data to be obtained from the \textit{IXPE} observations of an AMP. 
We substituted the source with the model described above and obtained the event list for a specified observing time of 600~ks for a source with a flux of 100~mCrab. In this part of our study, we do not investigate the energy dependency of PD, and collect all the data in one energy bin. 
These data sets are in exactly the same form as publicly available \textit{IXPE} data products, so we will be able to apply the developed routines to the real data as soon as they become available. 

\begin{figure*} 
  \centering
  \includegraphics[width=0.32\textwidth]{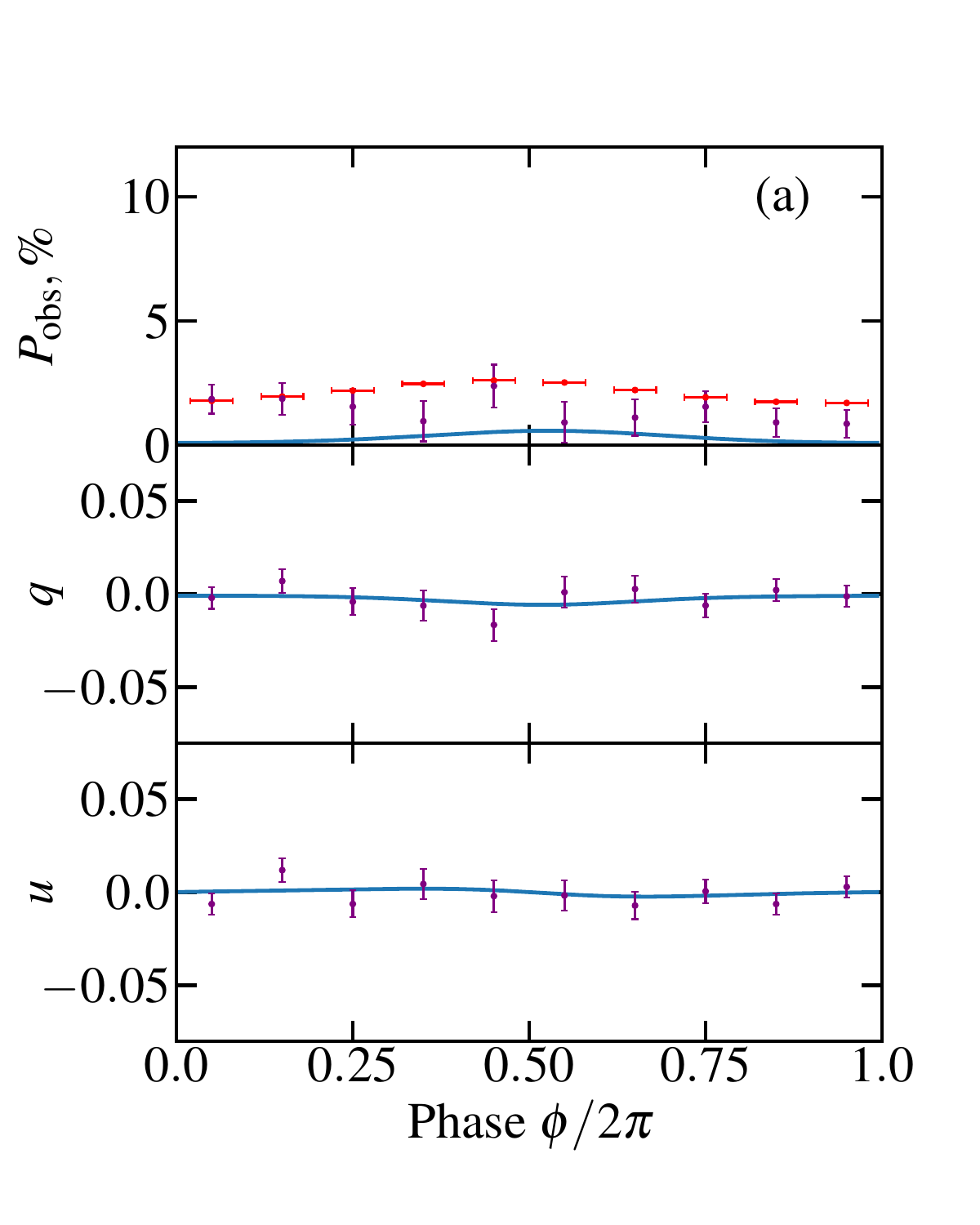}
  \includegraphics[width=0.32\textwidth]{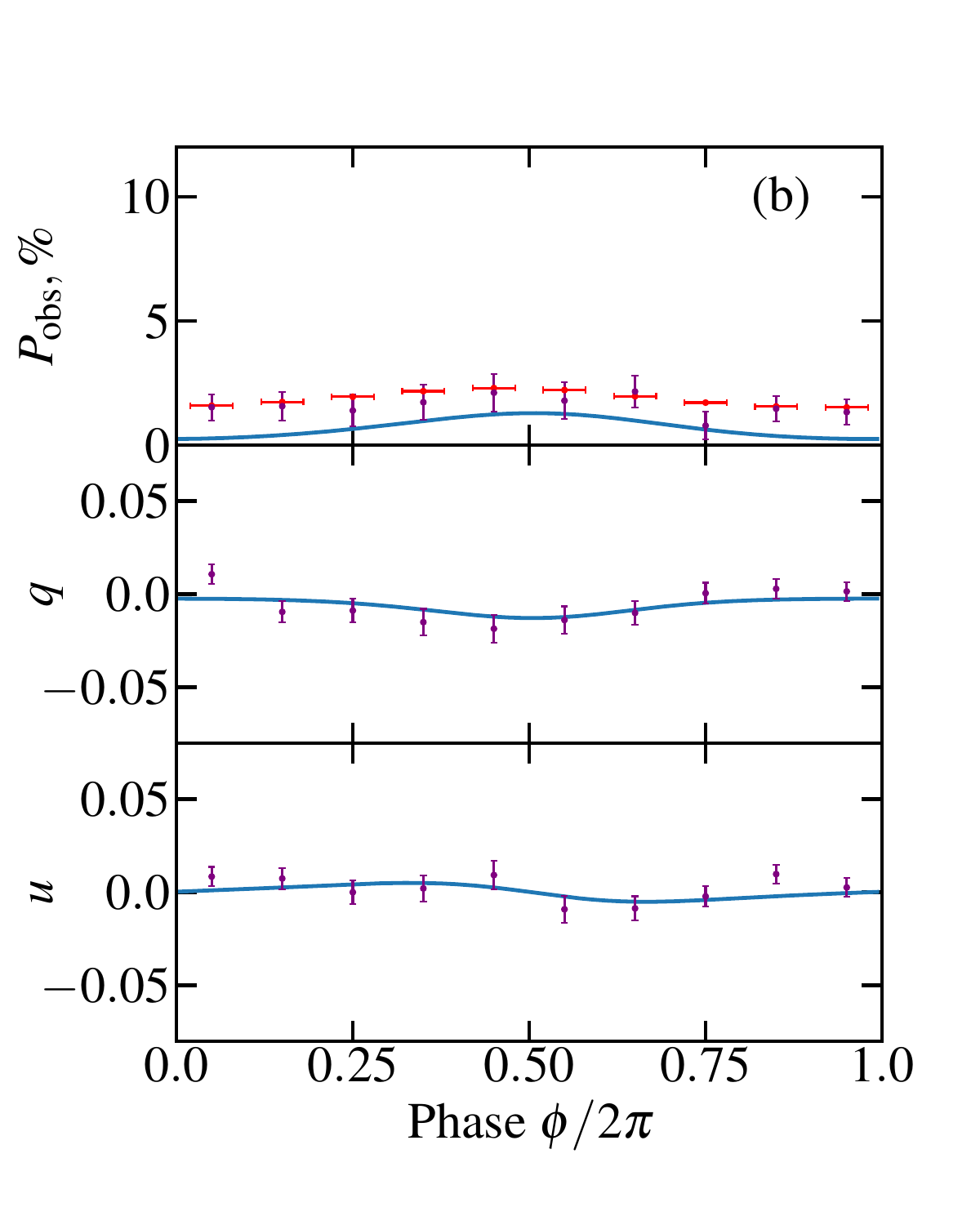}
  \includegraphics[width=0.32\textwidth]{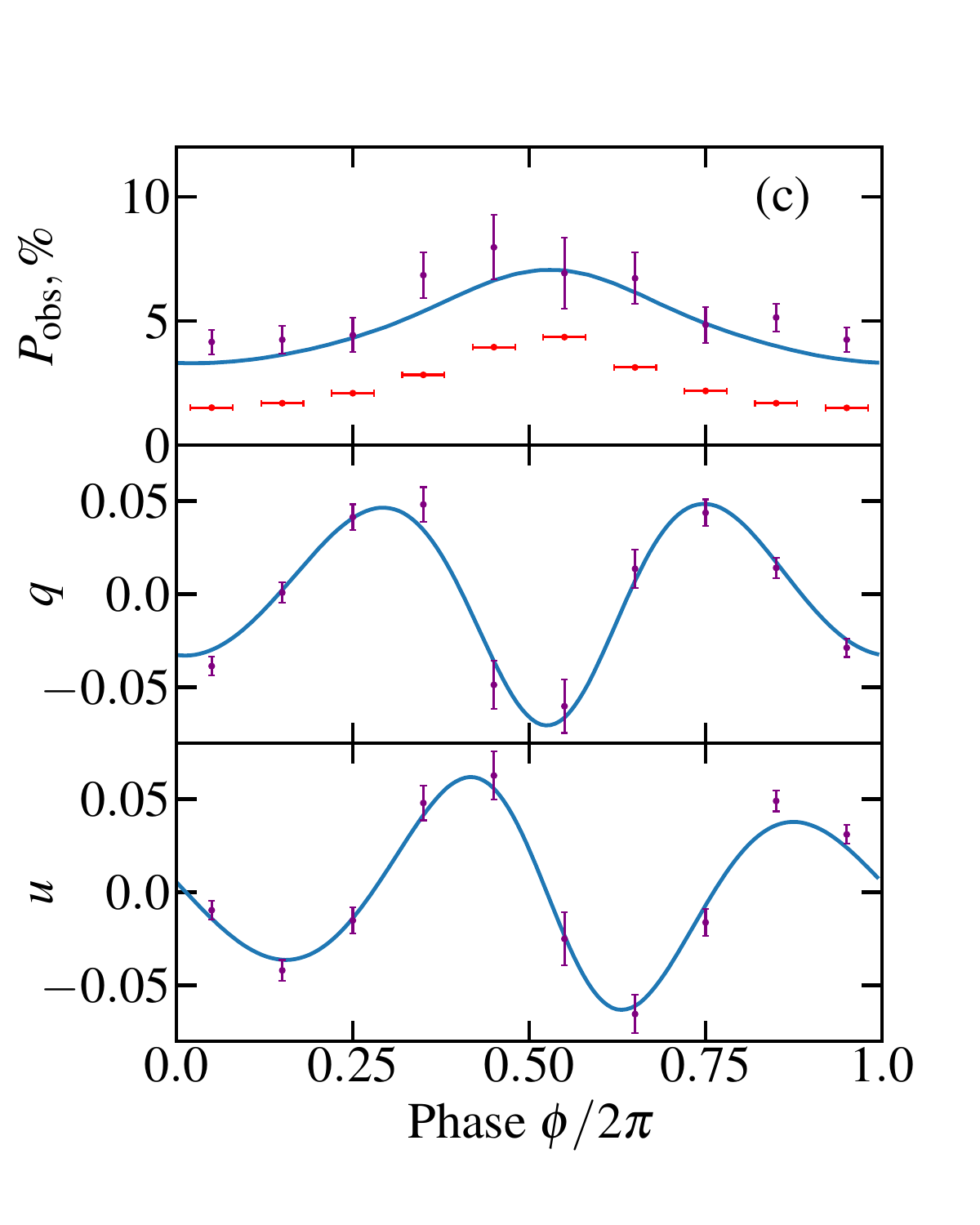}
  \caption{Simulated PD and normalized Stokes $q$ and $u$ profiles for the NS parameter set 1 (panels (a) and (b) are for the 2--8 keV and 2--5 keV energy range, respectively) and set~4 (c) from Table~\ref{fit_results}. The exposure time is 600~ks in all three cases. The blue lines show the theoretical model, the purple dots are the synthesized observed data, and the red dots are the MDP values for the corresponding phase bins.}
  \label{fig:model}
\end{figure*}

As mentioned above, applying the Compton scattering formalism reduces the expected PD as compared to the previous studies where Thomson scattering in the NS atmosphere was considered. 
We synthesized the data for the same parameter set as used in Fig.~\ref{fig:polpulse1} (see Table~\ref{fit_results}, Model~1) and compared the PD to the so-called minimal detectable polarization (MDP) at 99\% significance (see \citealt{WEO10} for a definition of the MDP and \citealt{KCB15} for further details) values for the corresponding phase bins. Figure~\ref{fig:model}a illustrates that the obtained values of the PD are below the MDP. According to \citet{WEO10}, this means that we cannot have a proper measurement of the PD in this case. 
From this point, there are three main ways to improve the quality of the potential measurement: (i) we can increase the number of counts to reduce the MDP values, (ii) study different energy bands to see if the PD values are higher there, and/or (iii) choose a more promising source, which in terms of simulated data means selecting different source parameters.

\begin{table*}[!ht]
    \centering
    \caption{Parameter sets for all the computed models and the most probable values obtained in the fitting.}
    \begin{tabular}{cccccccccccc}
    \hline\hline \\
    \multicolumn{1}{c}{\multirow{2}{*}{Parameter}} & \multicolumn{4}{c}{Injected values} & & \multicolumn{6}{c}{Fitting results} \\  \cline{2-5} \cline{7-12} 
    \\ \multicolumn{1}{c}{} & 1&   2 & 3 &  4 &  & 1a\tablefootmark{a} &  1b\tablefootmark{b} &  2 & 3a\tablefootmark{a} &   3b\tablefootmark{b} &  4  \\ \hline
        \\$i$ [deg] & 60 & 10 & 60 & 10 & & $25_{-12}^{+17}$ & $40_{-11}^{+18}$ & $10_{-3}^{+4}$ & $49_{-12}^{+15}$ & $54_{-10}^{+15}$ & $9_{-3}^{+3}$  \\[5pt] 
        $\theta$ [deg] & 20 & 105 & 20 &  105 && $21_{-12}^{+16}$ & $15_{-5}^{+7}$ & $101_{-6}^{+6}$ & $20_{-6}^{+7}$ & $17_{-3}^{+4}$ & $104_{-6}^{+4}$  \\[5pt] 
        $\chi$[deg] &0 & 0 & 0 & 0 & & $-2_{-30}^{+33}$ & $-7_{-9}^{+9}$ & $-6_{-14}^{+14}$ & $2_{-41}^{+8}$ & $-3_{-3}^{+3}$ & $-7_{-11}^{+11}$  \\[5pt]
        $\Delta \phi/2\pi$  &0.0 & 0.0 & 0.0 & 0.0 & & $0.01_{-0.24}^{+0.22}$ & $-0.02_{-0.11}^{+0.13}$ & $-0.02_{-0.04}^{+0.04}$ & $0.03_{-0.05}^{+0.05}$ & $-0.02_{-0.03}^{+0.03}$ & $-0.02_{-0.03}^{+0.03}$  \\[5pt]
        $T_{\mathrm{e}}$ [keV]  &50 & 50 & 50 & 50 && $58_{-24}^{+27}$ & $62_{-26}^{+26}$ & $59_{-25}^{+28}$ & $63_{-27}^{+26}$ & $65_{-27}^{+24}$ & $64_{-26}^{+24}$ \\[5pt] 
        $T_{\mathrm{bb}}$ [keV]  & 1.0 & 1.0 & 1.0 & 1.0 && $1.0_{-0.3}^{+0.3}$ & $1.0_{-0.3}^{+0.3}$ & $0.9_{-0.3}^{+0.4}$ & $1.0_{-0.3}^{+0.3}$ & $1.1_{-0.3}^{+0.3}$ & $0.9_{-0.3}^{+0.4}$  \\[5pt]
         $\tau_{\mathrm{T}}$  &1.0 & 1.0 & 1.6 & 1.6 && $1.4_{-0.6}^{+1.1}$ & $1.9_{-0.8}^{+1.0}$ & $1.6_{-0.6}^{+0.9}$ & $2.0_{-0.8}^{+0.9}$ & $2.2_{-0.9}^{+0.8}$ & $2.1_{-0.7}^{+0.8}$ \\[5pt] \hline
    \end{tabular}
    \tablefoot{
   NS mass, radius, and spin frequency are fixed at $M = 1.4 ~\msun$, $R$ = 12 km, $\nu$ = 401 Hz. These parameters, as well as Model 1 parameters, are selected based on the SAX J1808.4$-$3658 pulsar and expected values for canonical NSs. The errors represent the 68\% credible interval. Additional parameters $\chi$ and $\Delta \phi$ are the position angle of the pulsar rotation axis and the phase shift, respectively. As before, $i$ is  inclination, $\theta$ is co-latitude of the spot, $T_\mathrm{e}$ and $T_\mathrm{bb}$ are the temperatures of the electron in the slab and seed photons, respectively, and $\tau_\mathrm{T}$ is the Thomson optical depth of the slab. Values are averaged over two data realizations for each scenario. 
    \tablefoottext{a}{Columns labeled 1a and 3a correspond to fitting the whole energy band of \textit{IXPE}, 2--8 keV.} 
    \tablefoottext{b}{Columns labeled 1b and 3b correspond to the data realization generated and fitted in the 2--5 keV energy band. }
    }
    \label{fit_results}
\end{table*}

Increasing the number of counts by prolonging the exposure time is impossible with AMPs, as they are transient sources observable only during the outbursts, which typically last for a few weeks \citep[see e.g.,][]{Patruno2021}. 
Also, observations lasting  longer than 600 ks are unrealistic for scheduling reasons.  
The two other options are explored below. 

Studying different energy bands is relevant as, according to our predictions from Sect.~\ref{sec:emission_model}, the PD changes sign within the 2--8 energy band. This means that averaging over all energies decreases the total PD value. Figure~\ref{fig:model}b illustrates the results of analyzing the 2--5 keV energy band instead of the 2--8 keV represented in Fig.~\ref{fig:model}a. 
The blue line corresponding to the model prediction is now significantly higher and closer to the MDP values. However, the improvement is slight, and we note that the PD values are still below the MDP values in this case.

 \begin{figure*}
  \centering  \includegraphics[width=0.49\textwidth]{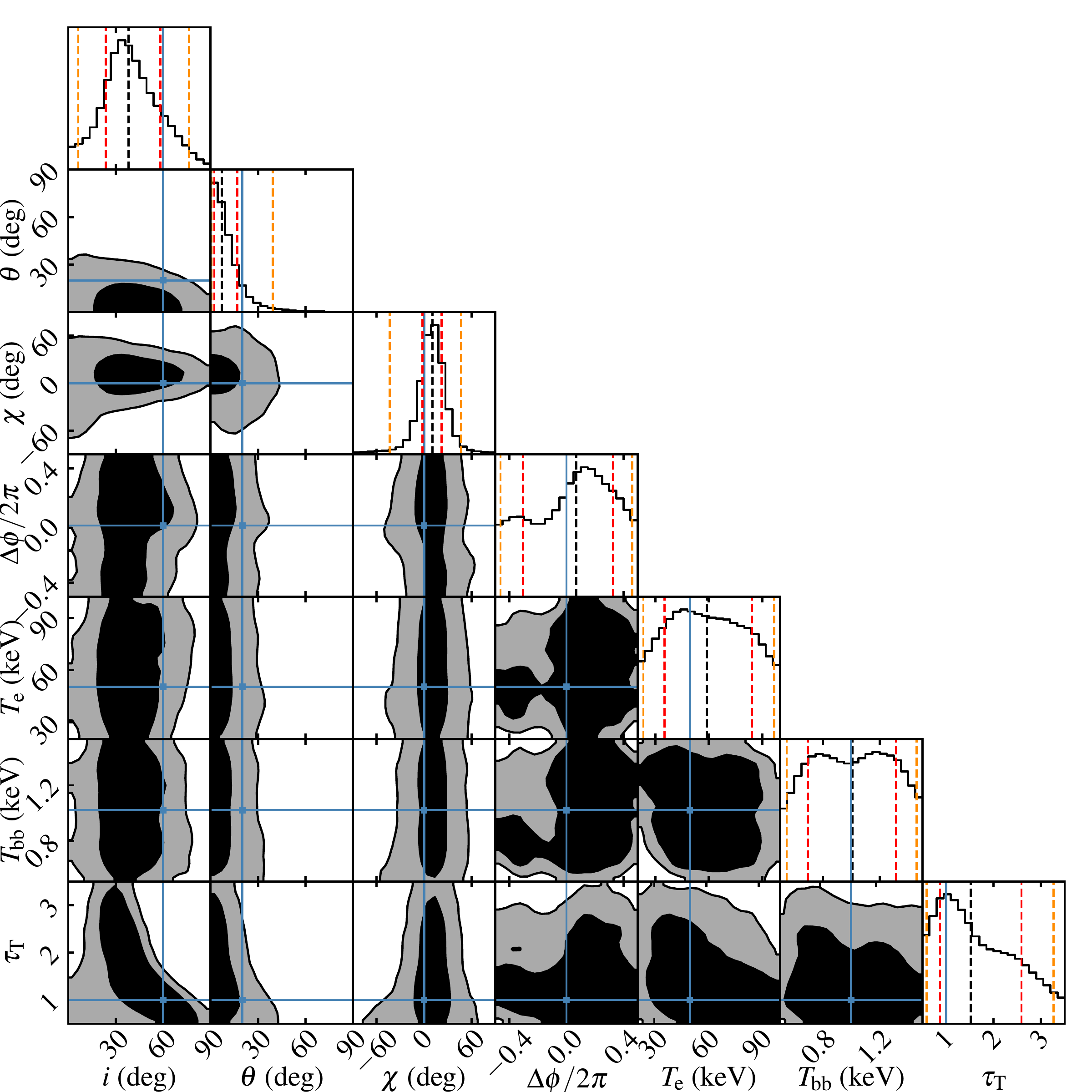}
  \includegraphics[width=0.49\textwidth]{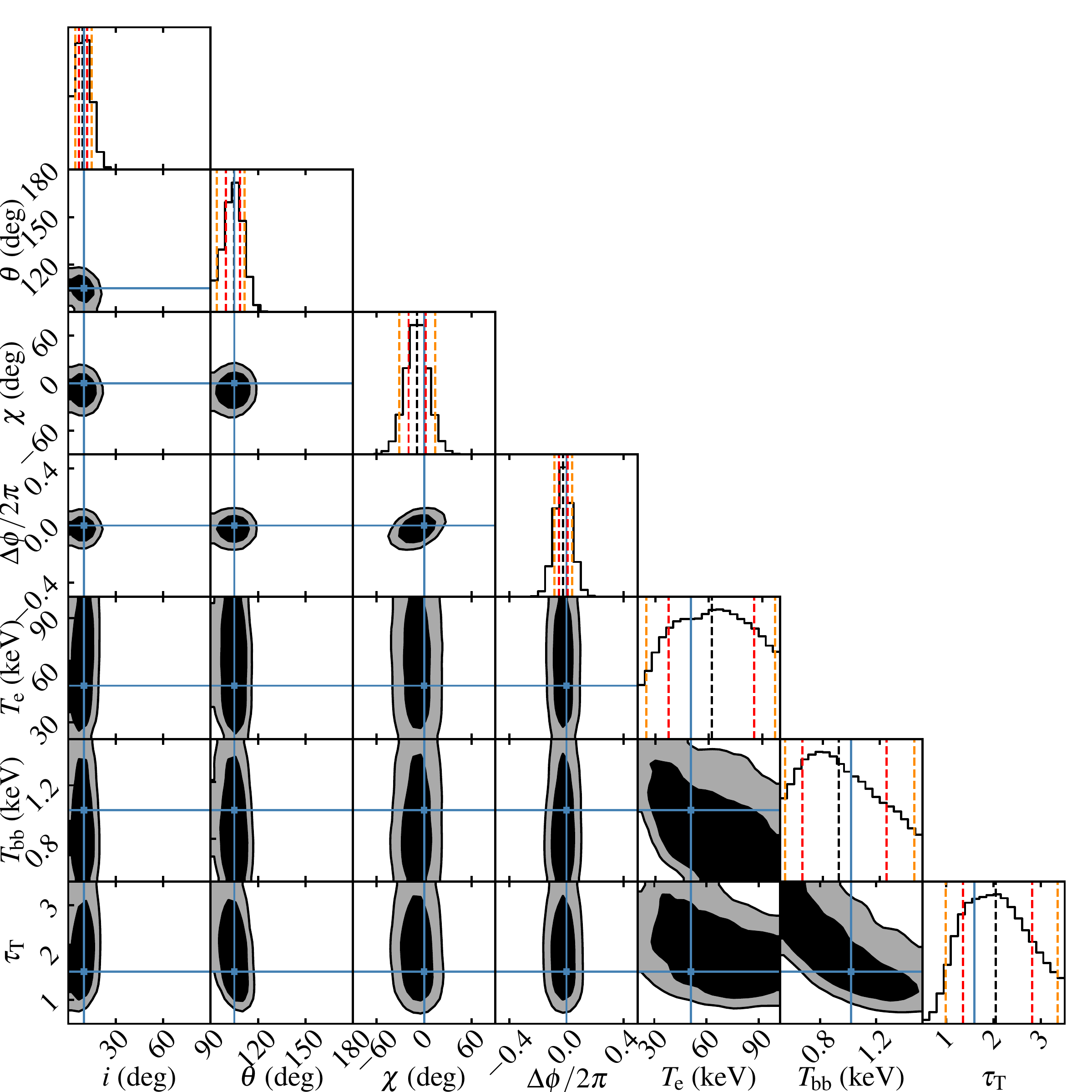}
\caption{Posterior probability distributions for NS parameters for Model 1a ($\textit{left panels}$) and Model 4 ($\textit{right panels}$) from Table~\ref{fit_results} using one data realization. 
The two-dimensional contours correspond to 68\% and 95\% credible regions with the corresponding regions colored black and gray. 
The histograms show the normalized one-dimensional distribution for a given parameter derived from the posterior samples. The mean value and the 68\% and 95\% credible levels for the  parameters are shown by the vertical dashed black, red, and orange lines, respectively.
Blue lines indicate the true values of the parameters that were used for the data simulations. }
  \label{fig:posterior}
\end{figure*}

In our search for a more optimistic (from the point of view of potential observations) yet still valid set of both geometrical and atmosphere parameters, we aimed to reproduce the known spectra of the AMPs with the parameters that, coming from our previous investigations, are expected to give a higher PD of the emission. 
From Fig.~\ref{fig:pmu}, we see that the optical depth affects the PD strongest from all the atmosphere parameters, and so we aimed to increase it but preserve the overall spectrum behavior presented in \citet{PG03}. 
We also attempted to increase the electron temperature to match the predictions in \cite{PG03}, but the scenario proposed there gives PD values below the MDP values, and so we chose not to continue with it. 
We developed two possible adjustments to the fiducial scenario (Model~1): we can choose the geometry such that the angle between the inclination and co-latitude of the spot is almost 90\degr\ (as in Model~2) or use `more promising' atmosphere parameters (as in Model~3). 
We can also combine these two (Model~4). The parameter values for these four models are given in the left part of Table~\ref{fit_results} in the columns labeled correspondingly 1--4. 
Figure~\ref{fig:model}c shows that for Model~4, the PD now exceeds the MDP values. We also checked this for Models~2 and 3, and for Model~2 the PD is slightly higher than the MDP values, while for Model~3 the PD is similar to MDP. 
In all the models, we only calculated the emission coming from one spot on the surface of the NS. 

In addition, we attempted to improve the quality of the results for Model 1 illustrated in Fig.~\ref{fig:model}a,b by reducing the number of phase bins, as the PD varies less in these cases compared to other models. We find that for Model 1, case a, only reducing the number of bins to one gives us the PD value that barely exceeds the MDP, meaning that we can detect something from such a source, but we would lose the phase variability completely. For Model 1, case b, we can obtain a detection with three phase bins, but only for the bin that includes the peak (in Fig.~\ref{fig:model}b the corresponding phase shift is 0.5), and again, the PD barely exceeds the MDP. We conclude that reducing the number of bins can support the observations with barely measurable PD, but its impact is not sufficient for us to rely upon it in cases where the source is otherwise unobservable with \textit{IXPE}.  

Nevertheless, the most recent development of the \textit{IXPE} data analysis shows that for multiple-point observations (i.e., observations split in different phase bins), information about polarization can be extracted even from the data points where PD values do not exceed MDP \citep{GCH23, Suleimanov23}. As \textit{IXPE} measures the Stokes parameters $q$ and $u$ rather than PD, we can fit these parameters directly and obtain some constraints on the geometry of the source. This approach is further explored in the following section.

\subsection{Fitting the data}

After synthesizing the data, we applied the \textsc{multinest} ~\citep{multinest09} multimodal nested sampling algorithm (specifically, the \textsc{pymultinest} package, see \citealt{pymultinest}) to obtain new constraints on both the geometrical and atmosphere parameters of the source. 
As in ~\citet{SLK21}, we fit the geometrical parameters, such as inclination, co-latitude, phase shift, and position angle, but unlike there, we also perform a simultaneous fit over the parameters of the atmosphere, such as characteristic photon temperature, electron temperature, and optical depth of the thermalization layer. Table~\ref{fit_results} presents the most probable values and 68\% credible intervals for these parameters. 
Figure~\ref{fig:posterior} shows the results of the fitting presented for the fiducial set of parameters (Model 1) and for a `more promising' scenario (Model 4). In both cases, the full energy band of \textit{IXPE} (2--8 keV) is considered, and the exposure time is 600 ks. 

Figure~\ref{fig:posterior} clearly illustrates the improvement of the fitting results if comparing the data synthesized from the Model 1-type of source (left) and the data coming from the Model 4-type of source (right). As we use the uniform prior for all our fittings, in the former case, the constraints originate from our direct fitting of the $q$ and $u$ Stokes parameters. This illustrates the possibility to extract some information even when the MDP values are not reached. In the latter case, we can clearly see the significant constraints coming from the PD values exceeding the MDP by approximately a factor of two.

We can also conclude that the constraints on the atmospheric parameters are notably less restrictive than the ones on the geometrical parameters of the NS. The main reason for obtaining such broad constraints on atmospheric parameters lies in the energy range of the \textit{IXPE} satellite, as we are unable to detect the whole spectrum and fit it correctly. A solution to this issue could be to use the observational results from several missions simultaneously; for instance, fitting the \textit{NuSTAR} data in its higher energy range to determine the electron temperature (and consequently the optical depth through the spectral slope). Then, using the values obtained, we can fit the \textit{IXPE} data with the fixed (or constrained) atmospheric parameters. However, we see that even with large uncertainties on these parameters, we are able to derive constraints on the geometry of the source from the \textit{IXPE} data only. 

Table~\ref{fit_results} also shows that the constraints from the Model~3 data are less restrictive than the ones from Model~2, and so we can conclude that the atmospheric parameters are less important in our search for the `promising' source than the geometric parameters of the NS. 
We also attempted to improve the constraints by reducing the number of simultaneously fitted parameters (e.g., we fixed the atmospheric parameters and only fitted the geometrical parameters). Decreasing the number of fitted parameters reduces the calculation time, but the credible intervals for the fitted parameters remain the same. 

In the fittings presented here, we fixed the mass and radius of the NS. This is done for the simplicity of the calculations, as they are quite time-consuming. Moreover, we only fit the normalized Stokes $q$ and $u$ parameters, and they are not particularly sensitive to the mass or radius. We could also fit the Stokes $I$ parameter (i.e., flux) as well and add mass and radius as free parameters. Adding two more parameters would increase the computational time significantly, but we do not expect any significant impact of this on the credible intervals of other parameters. 

\section{Discussion}\label{sec:discussion}
Our fiducial set of NS parameters is based on the SAX J1808.4$-$3658, and Fig.~\ref{fig:polpulse1} shows that, for the parameters selected, our predictions for the PD of the SAX J1808.4$-$3658-like source are low. However, with the uncertainties we still have on  both the geometry and the atmospheric parameters of the source, there is still a possibility to observe a higher PD. 
Furthermore, with about 20 AMPs discovered, we can put forward several conclusions about the perspective of observing measurable PD from this class of sources. 

It is known that the product of the electron temperature and optical depth for the AMPs is almost invariant \citep{Poutanen:2005qx}. Nevertheless, we have learned that all the atmospheric parameters have their own contribution to the PD values; see Fig.~\ref{fig:pmu}. Optical depth is the most important parameter here; we are searching for a source with an atmosphere that is sufficiently optically thick for many Compton scattering events to occur, yet not so thick that the pulsation pattern is excessively influenced by these scatterings. In this sense, objects such as IGR~J17591$-$2342 and  IGR~J17511$-$3057 are promising. The former has an estimated optical depth of 1.59--2.3 and a pulsed fraction of the emission of 10\%--17\% \citep{KuiperTsygankov, MancaSanna2018}. The latter has a lower optical depth of 1.34, but higher electron and seed photons temperatures ($T_\mathrm{e}=51$~keV, $T_\mathrm{bb}=1.36$~keV) and a pulsed fraction of 14\% \citep{Papitto10}. 
Constraints on the geometrical parameters of the AMPs are among the goals of the polarimetric observations. So far our main source of information on the possible geometry is the amplitude of the pulsation, which depends approximately on the product of the sines of the inclination and the magnetic co-latitude \citep{PB06}. Moreover, while for the inclination we usually have some constraints, little is known about the geometrical configuration of the magnetic field of the AMP. From the emission patterns shown in Fig.~\ref{fig:Compton1}, the highest PD is received when the  line of sight and the normal to the emitting area are orthogonal. One of the most interesting objects to study is Swift J1749.4$-$2807, for which the inclination is well-constrained, $74.4\degree<i<77.3\degree$, and the co-latitude is estimated to be $\theta \approx 50\degree$ \citep{Altamirano10}. 
In this case, we can expect high polarization and we can test whether the inclination obtained from the polarimetry actually matches the orbital one. 

We performed all our fittings with 600 ks of exposure time. It is typical for AMPs to go into outburst for a couple of weeks, and so in order to collect enough photons we need either a rather soft spectrum of emission or a very bright source. We can exclude hard sources such as IGR J18245$-$2452 with a photon index of $\Gamma \approx 1.4$ \citep{Papitto13}.  

Figure~\ref{fig:polpulse1} illustrates an additional feature of the PD behavior we expect to observe: the observed PD is different between cases where the secondary spot is invisible (e.g., the very beginning of the outburst) and where it is visible (at the later stages of the outburst). In order to get additional information about the geometry of the source, it is crucial to observe the outburst from the earliest  to the latest possible stages. 

We purposely focused on phase-resolved analysis and did not investigate the possibilities that energy-resolved analysis could open up. \textit{IXPE} has an energy resolution of 0.57 keV (at 2 keV, \citealt{IXPE_tech_4}), and so with sufficient statistics it is possible to study the behavior of PD with energy and obtain more information about the geometry and atmospheric properties of the source. 

The model presented in this paper can be improved further. For instance, the isothermal (static) slab approximation of the accretion shock can be replaced with a more physical model where the dynamics of the accreting gas is computed self-consistently with the electron temperature. 
Also, the model for seed soft photons can be modified to account more accurately for the reprocessing of the hard X-rays at the NS surface, accounting for the polarization properties of this radiation. 
We intend to work on these developments in the future. 

\section{Summary}\label{sec:summary}

We present a new model to compute the flux and polarization of the emission coming from the accretion-powered millisecond pulsars. We used the formalism for Compton scattering in a  plane-parallel isothermal hot slab to describe the propagation of photons in the accretion shock above the NS surface. Figures~\ref{fig:Compton1} and \ref{fig:Compton2} show the result of such calculations for the fiducial set of parameters taken from \citet{SLK21}. Comparing our results with the previous studies using Thomson scattering formalism shows that, at least for the same NS parameters, our model predicts lower PD values. 

We calculated the Stokes parameters of the radiation as a function of two variables, the zenith angle of the emitted photon and its energy, for almost 14,000 combinations of three parameters: electron temperature, the temperature of the seed blackbody photons, and the Thomson optical depth of the slab. The tables of the spectral energy distribution of the Stokes parameters for all the parameters can now be used for further analysis of the accretion-powered millisecond pulsars, both in studying the pulse profiles and the polarimetric properties of the emerged radiation. Figure~\ref{fig:pmu} illustrates one of the ways to use this new data set. We can understand the impact of the NS atmosphere parameters on the angular dependency of the PD from this particular study. 

We then used the relativistic rotating vector model for the oblate NS to produce observed Stokes parameters as a function of the pulsar phase. Figure~\ref{fig:polpulse1} shows the observed fluxes coming from the two antipodal spots of an AMP and the combined total flux as calculated using our model and a total flux calculated using a previously developed Thomson model. A significant drop in the PD between these two models is again present here. 

We used the developed model of the AMP emission to generate the synthetic data sets imitating the one that the \textit{IXPE} satellite could produce after observing an AMP. We note that the NS with the parameters from our fiducial set would emit the light with PD values below the MDP values of \textit{IXPE} (which still does not exclude the possibility to extract information about the geometry of the source); however, for a different set of NS parameters, the PD is high enough to be detectable by \textit{IXPE}, as shown in Fig.~\ref{fig:model}. 
We then applied the multimodal nested sampling technique to fit the simulated data with our model and obtain  constraints on both the geometrical and atmospheric parameters of the NS. The results of the fitting are presented in Fig.~\ref{fig:posterior}. The main conclusion of this study is that we can obtain  reasonably good constraints on the geometrical parameters from the \textit{IXPE} data for some of the AMPs; however, the parameters of the atmosphere of the NS need to be constrained from the supporting observations by other X-ray missions such as \textit{NuSTAR}. 

Our findings contribute to the advancement of understanding AMPs and have significant implications for future research in this field. We are waiting for the AMPs to be observed by \textit{IXPE} so that we may apply the developed formalism to analyze the resulting data. We also hope to see the \textit{eXTP} mission observing the AMPs, as it would solve several of the issues discussed in this article: the larger effective area will lower the MDP, and so the fainter sources will become observable; the broader energy range will allow us to develop better constraints on the atmospheric parameters; and finally, the wide field instrument of \textit{eXTP} will be able to catch the transients, such as AMPs, and observe the outbursts from their very beginning. 

\begin{acknowledgements}
This research was supported by the Academy of Finland grant 333112 and by the Finnish Cultural Foundation grant 00220175 (AB).
TS acknowledges support from ERC Consolidator Grant (CoG) No. 865768 AEONS (PI: A.~Watts).
The computer resources of the Finnish IT Center for Science (CSC) and the FGCI project (Finland) are acknowledged. 
We thank Anna Watts, Bas Dorsman, Alessandro Di Marco, and John Rankin for discussions and testing our models.
\end{acknowledgements}

\bibliographystyle{aa}
\bibliography{main}

\end{document}